 \documentclass[pra,10pt,showpacs]{revtex4}
 \usepackage{graphicx}
\begin{document}

 \def\BE{\begin{equation}}
 \def\EE{\end{equation}}
 \def\BA{\begin{array}}
 \def\EA{\end{array}}
 \def\BEA{\begin{eqnarray}}
 \def\EEA{\end{eqnarray}}
 \def\nn{\nonumber}
 \def\ra{\rangle}
 \def\la{\langle}
 \def\E{{\cal E}}
 \def\tE{\tilde{\cal E}}
 \def\g{\gamma_\perp}
 \def\k{\kappa}
 \def\T{{\cal T}}

 \title{Cavity-assisted atomic Raman memories beyond the bad cavity limit: effect of four-wave mixing}
 \author{N.~G.~Veselkova}
 \author{N.~I.~Masalaeva}
 \author{I.~V.~Sokolov}
 \email{i.sokolov@mail.spbu.ru, sokolov.i.v@gmail.com}
 \affiliation{St. Petersburg State University,  7/9 Universitetskaya nab., St. Petersburg,
 199034 Russia}
 %\date{}

 \begin{abstract}

Quantum memories can be used not only for the storage of quantum information, but also for
substantial manipulation of ensembles of quantum states. Therefore, the speed of such manipulation and the
ability to write and retrieve the signals of relatively short duration becomes important. Previously there have been
considered the limits on efficiency of the cavity-enhanced atomic Raman memories for the signals whose
duration is not much larger than the cavity field lifetime, that is, beyond the bad cavity limit.
We investigate in this work the four-wave mixing noise that arises by the retrieval of the relatively short signals from
the cavity-assisted memories, thus complementing recent considerations by other authors, who mainly concentrated
on the limit of large cavity decay rate. The four-wave mixing noise is commonly recognized as an important factor,
able to prevent achieving a high memories quality in a variety of the atomic, solid state etc. implementations.

The side-band noise sources (with respect to the quantized signal, supported by the cavity) play important role in
the four-wave mixing. We propose an approach that allows one to account for the side-band
quantum noise sources of different physical origin in the cavity-assisted atomic memories using a unified theoretical
framework, based on a two-band spectral filtering of the noise sources. We demonstrate that
in such spectrally-selective memories the side-band atomic noise sources essentially
contribute to the four-wave mixing noise of the retrieved signal on a par with the side-band quantized field entering
the cavity.

 \end{abstract}


 \maketitle

 %\tableofcontents

 \section{Introduction}

Efficient quantum memories for light  \cite{Hammerer10} -- \cite{Heshami16} are viewed at as an
important component of many schemes of quantum information, such as quantum repeaters, quantum networks,
quantum computers etc. Of particular interest for future applications are the schemes that allow for the storage
and manipulation of the signals with many spatial and (or) temporal degrees of freedom.
In the atomic memories, exploiting cold  atomic ensembles as the storage medium, the resources for the essentially
multimode operation are provided by the independent spatial waves (quantum holograms) of the collective spin excitation
\cite{Vasilyev08} -- \cite{Vetlugin16} and by the time multiplexing \cite{Nisbet13}.

The cavity-enhanced atomic memories implemented experimentally with the use of cold \cite{Nisbet13,Bao12,Bimbard14}
and warm \cite{Saunders16} atomic ensembles demonstrate good efficiency and fidelity of quantum state manipulation.
The cavity enhances the coupling between the signal field and the storage medium by means of multiple passes
of light through the atomic ensemble, thus increasing the cooperativity parameter and the cavity field
lifetime. On the other hand, a co-processing in the memory of a time sequence of quantized
signals \cite{Vetlugin16,Nisbet13} within the time interval of effective storage implies shortening of the signals
duration. In view of this, it is natural to address a question: to what extent one can speed-up the manipulation of
a signal in the sequence, achieving maximal information content for the whole ensemble of signals. The
theoretical estimates \cite{Gorshkov07,Stanojevic11}, performed mostly in the bad cavity approximation, have
revealed that the memories quantum efficiency close to unity is achievable in this limit.

Recently we have investigated \cite{Veselkova17a} the limits on quantum efficiency of the cavity-enhanced atomic
Raman memories for the signals whose duration is not much larger than the cavity field lifetime, that is,
beyond the bad cavity approximation. There was determined \cite{Veselkova17b}
the needed non-stationary amplitude and phase behavior of strong classical control field that matches the
desirable time profile of both the envelope and the phase of the retrieved quantized signal.

The four-wave mixing noise present in real schemes of the atom-field interaction in atomic memories is commonly
recognized as an important factor, able to prevent achieving a high memories quality.
The four-wave mixing arises when besides the memory channel $\Lambda$--scheme
there is involved an additional $\Lambda$--scheme. In this $\Lambda$--scheme the same control field produces, via the Raman
two-quantum transition, the pairs of quanta: the quantized field excitation, and the collective spin wave excitation
(the spin polariton), in analogy to parametric scattering in presence of the $\chi^{(2)}$ nonlinearity.
The arising spin excitations are involved in the readout  process on a par with the stored signal.

The deteriorative effect of four-wave mixing noise on the atomic memories overall efficiency was investigated both
for the single-pass \cite{Lauk13,Geng14,Prajapati17} and the cavity-assisted \cite{Saunders16,Nunn17} schemes.
Prajapati et al. \cite{Prajapati17} proposed to suppress the four-wave mixing in the single-pass configuration by
introducing two-quantum Raman absorption channel for the side-band light. In the cavity-assisted atomic memories,
the spectral filtering of the off-resonant side-band quantized field, performed by the cavity, also makes it possible
to suppress effectively the four-wave mixing noise \cite{Nunn17}. The approach exhibited in \cite{Nunn17} is based
on the explicit description of the quantized side-band noise field as an independent wave, performing round-trips inside the
cavity. The detailed theoretical analysis  \cite{Nunn17} of the four-wave mixing noise is focused mainly on the atomic
memory operation in the bad cavity limit.

We present in our work a theoretical research of the four-wave mixing noise in the cavity-assisted atomic Raman
memories, valid also for the signals whose duration is not much larger than the cavity field lifetime, that is, beyond
the bad cavity limit. Our generalization may occur potentially useful for the analysis of the essentially multimode
regimes of the memories operation.

We account for the side-band quantum noise sources of different physical origin using a unified theoretical
framework and do not restrict ourselves by considering only the noise due to the side-band noise light field.
We demonstrate that in the spectrally selective cavity-assisted atomic memories the side-band atomic noise
sources essentially contribute to the four-wave mixing noise of the retrieved signal on a par with the side-band
field entering the cavity.

The finally estimated quantity is the noise variance of quadrature amplitudes of the output signal, observed
by means of an optimal homodyne detection. We present the noise contributions, associated with the four-wave mixing
and with the different from unity quantum efficiency, for a wide range of the signal duration, including
the signals whose duration does not much exceed the cavity field lifetime.

%%%%%%%%%%%%%%%%%%%%%%%%%%%%%%%%%%%%%%%%%%%%%%%%%
%%%%%%%%%%%%%%%%%%%%%%%%%%%%%%%%%%%%%%%%%%%%%%%%%
%%%%%%%%%%%%%%%%%%%%%%%%%%%%%%%%%%%%%%%%%%%%%%%%%
%%%%%%%%%%%%%%%%%%%%%%%%%%%%%%%%%%%%%%%%%%%%%%%%%

 \section{Memory cell in presence of four-wave mixing}
 \label{sec_Memory_cell}

 \begin{figure}
 \begin{center}
 \includegraphics[width=0.6\linewidth]{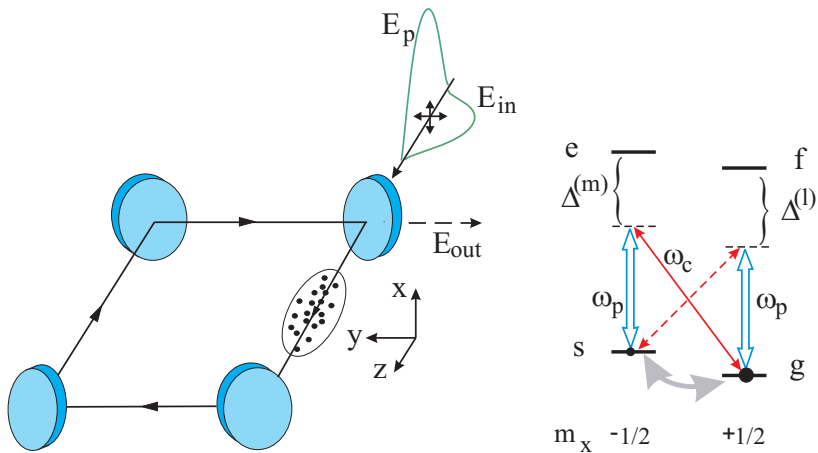}
 \caption{The memory schematic and the relevant Raman transitions in atoms.}
 \label{fig_1}
 \end{center}
 \end{figure}
The memory scheme to be considered appears in figure \ref{fig_1}.
The Hamiltonian of the electric dipole interaction of $N$ motionless atoms with the cavity field in the rotating wave
approximation is given by
 $$
 H = H_0 + V,
 $$
 $$
 H_0 = \hbar \omega_{c}a^{\dagger}a + \hbar\sum_{j=1}^N
\left(\omega_{sg}\sigma^{(j)}_{ss} + \omega_{eg}\sigma^{(j)}_{ee} +
\omega_{fg}\sigma^{(j)}_{ff}\right),
 $$
 $$
 V = - \hbar \sum_{j=1}^N \left[\Omega^{(m)}(t)\sigma^{(j)}_{es}
e^{-i\omega_{p}t} + \Omega^{(l)} (t)\sigma^{(j)}_{fg}
e^{-i\omega_{p}t} + a g^{(m)}\sigma^{(j)}_{eg} +
a g^{(l)}\sigma^{(j)}_{fs}\right] + h.c.
 $$
Here $a$ and $\sigma^{(j)}_{nm} = (|n\ra\la m|)^{(j)}$ are the quantized cavity field and the atomic transition operators
for j-th atom, $\omega_{nm}$ is the transition frequency, $\Omega^{(m)}(t)$ and $\Omega^{(l)}(t)$ are the Rabi
frequencies in the left (memory channel) and in the right  $\Lambda$-schemes. The latter one is responsible for the
control field Stokes Raman scattering, which results in the generation of bosonic quanta pairs (the light and the collective spin).
Having in mind an analogy to the parametric generation of pairs in $\chi^{(2)}$ non-linear media, we will call
the interaction channel, introduced by the right $\Lambda$--scheme, the Raman luminescence or, briefly,
the luminescence channel.

The cavity frequency $\omega_c$ and the classical control field frequency
$\omega_p = \omega_c - \omega_{sg}$ are matched so that to support the resonance condition for memory channel,
$g^{(m)}$ and $g^{(l)}$ are the coupling parameters for quantized mode field in the corresponding channel.

The single cavity mode approximation implies that the frequency mismatch $\sim|\omega_{sg}|$ between the memory
and the luminescence channels is small as compared to the frequency distance between the cavity modes.
The spatial factors are omitted in the Hamiltonian since for the co-propagating control and quantized fields the difference
between the longitu\-di\-nal wave numbers $k_{pz} - k_{cz}$ does not manifest itself on the atomic cloud length. That is,
we do not consider here the spatial addressability resource which allows for an essentially multimode memory operation
\cite{Vetlugin16}.

The slow amplitudes of the field and the collective atomic observables are introduced as
 \BE
 \label{slow_ini_1}
 {\cal E}(t) = a(t)\exp(i\omega_c t),
 \EE
 \BE
 \label{slow_ini_2}
 \sigma_{ge}(t) = \sum_{j=1}^N \sigma^{(j)}_{ge}(t) e^{i\omega_ct }, \quad
 \sigma_{gs}(t) = \sum_{j=1}^N \sigma^{(j)}_{gs}(t) e^{i\omega_{sg}t},
  \EE
 \BE
 \label{slow_ini_3}
\sigma_{sf}(t) = \sum_{j=1}^N \sigma^{(j)}_{sf}(t) e^{i\omega_c t}, \quad \sigma_{gf}(t) =
\sum_{j=1}^N \sigma^{(j)}_{gf}(t) e^{i(\omega_{c}+\omega_{sg})t}.
 \EE
%
%Note that the introduced above atomic coherence $\sigma_{gf}$ is defined as slow for the transition,
%where $\sigma_{gf}$ is coupled to $\sigma_{gs}$ by  the absorption of cavity photon.
The Heisenberg equations of motion are derived and linearized under the assumption of
almost unchanged initial population of the ground state g, $\sigma^{(j)}_{gg}(t)\to 1$, when the population of
all other states, as well as the cross-coherences $\sigma^{(j)}_{es}$ and $\sigma^{(j)}_{fe}$, can be neglected,
with the exception for the polarization $\sigma^{(j)}_{sf}$, which couples the cavity field to the luminescence
channel.

By introducing the cavity field decay at the rate $\kappa$ and the transverse relaxation at rate ${\g}$, induced by
the upper states decay, we arrive at
 \begin{equation}
 \label{dot_E_ini}
\dot{\cal E}(t) = -\kappa{\cal E}(t)+ig^{(m)}\sigma_{ge}(t)+
ig^{(l)}\sigma_{sf}(t)+ \sqrt{2\kappa}{\cal E}_{in}(t),
 \end{equation}
 \begin{equation}
 \label{dot_ge_ini}
\dot\sigma_{ge}(t) = -({\g}+ i\Delta^{(m)})\sigma_{ge}(t) +
i\Omega^{(m)}(t)\sigma_{gs}(t)
+ ig^{(m)}N{\cal E}(t) + \sqrt{2{\g}N}F_{ge}(t),
 \end{equation}
 \begin{equation}
 \label{dot_gs_ini}
 \dot\sigma_{gs}(t) = i\Omega^{(m)*}(t)\sigma_{ge}(t)
 -i\Omega^{(l)}(t)e^{i2\omega_{sg}t}\sigma_{fs}(t) + ig^{(l)}{\cal
 E}^{\dagger}(t)\sigma_{gf}(t),
 \end{equation}
 \begin{equation}
 \label{dot_gf_ini}
\dot{\sigma}_{gf}(t)= - (\g+i(\Delta^{(l)}- 2\omega_{sg}))\sigma_{gf}(t) +
i\Omega^{(l)}(t)Ne^{i2\omega_{sg}t} + ig^{(l)}{\cal E}(t)\sigma_{gs}(t)+
\sqrt{2{\g}N} F_{gf}(t),
\end{equation}
 \begin{equation}
 \label{dot_sf_ini}
\dot\sigma_{sf}(t) = - (\g+i(\Delta^{(l)}- 2\omega_{sg}))\sigma_{sf}(t) +
i\Omega^{(l)}(t)e^{i2\omega_{sg}t}\sigma_{sg}(t)+ \sqrt{2{\g}N}F_{sf}(t).
\end{equation}
The input cavity field ${\cal E}_{in}(t)$ and the Langevin noise operators $F_{nm}(t)$ (corresponding to
vacuum noise fields) satisfy the standard relations,
 \begin{equation}
 \label{noise_ini}
[{\cal E}_{in}(t),{\cal E}_{in}^\dagger(t')] = [F_{ge}(t),F^\dagger_{ge}(t')] =
[F_{gf}(t),F^\dagger_{gf}(t')] = \delta(t-t'),
 \end{equation}
which preserve the commutation relation for the cavity field, $[{\cal E}(t),{\cal E}^{\dagger}(t)]=1$,
and the properties of the atomic observables of the form $\sigma_{kl}(t)\sigma_{mn}(t) =
\delta_{lm}\sigma_{kn}(t)$ (the Einstein's theorem). The same assumption of the unchanged initial
population of the ground state g was used when deriving (\ref{noise_ini}). The noise source $F_{sf}$ has
zero power in this limit and will be omitted.

Now we introduce, in analogy to \cite{Veselkova17a,Veselkova17b}, physically reasonable corrections to the field and
atomic frequencies.  The cavity mode frequency shift due to the  linear refractive index is
 \BE
 \label{mode_shift}
 \delta_c = - \frac{g^{(m)2}N}{\Delta^{(m)}}.
 \EE
The dynamic correction $\delta_s(t)$ to the frequency $\omega_{sg}$ of s - g transition due to the AC Stark shifts,
induced by the strong control field (which might be not mutually compensating in general case), results in the additional
phase $\varphi_s(t)$ of the collective spin coherence, $\sigma_{gs}(t) \sim \exp(-i\varphi_s(t))$, where
 \BE
 \label{spin_shift}
 \delta_s(t)= -\left(\frac{|\Omega^{(m)}(t)|^2}{\Delta^{(m)}}-
 \frac{|\Omega^{(l)}(t)|^2}{\Delta^{(l)}}\right), \qquad
 \varphi_s(t) = \int_0^t dt' \delta_s(t').
 \EE
These phase corrections are incorporated into a new self-consistent set of the slow field and atomic variables,
 \BEA
 \E(t) = e^{-i\delta_c t} \tE(t), \quad  \E_{in}(t) = e^{-i\delta_c t} \tE_{in}(t),\\
 \sigma_{ge}(t) = e^{-i\delta_c t} \tilde\sigma_{ge}(t), \quad
 \sigma_{gs}(t) = e^{-i\varphi_s(t)} \tilde\sigma_{gs}(t),\\
 \sigma_{gf}(t) = e^{-i(\delta_c t + \varphi_s(t))} \tilde\sigma_{gf}(t), \quad
 \sigma_{sf}(t) = e^{-i\delta_c t} \tilde\sigma_{sf}(t),\\
 \Omega^{(m)}(t) = e^{-i(\delta_c t - \varphi_s(t))} \tilde\Omega^{(m)}(t), \quad
 \Omega^{(l)}(t) = e^{-i(\delta_c t - \varphi_s(t))} \tilde\Omega^{(l)}(t).
 \EEA
New noise operators  are defined similarly to the corresponding atomic variables.

Next, we substitute these definitions to the basic equations above  and perform the first adiabatic elimination.
That is, we assume the Raman regime condition when both frequency mismatches
$|\Delta^{(m)}|$ and $|\Delta^{(l)}|$ are much larger than other frequency parameters of the scheme.
The quantities (d/dt)$\tilde\sigma_{ge}$, (d/dt)$\tilde\sigma_{gf}$, and (d/dt)$\tilde\sigma_{sf}$ are set to zero,
the corresponding observables are expressed in terms of other variables and substituted into the remaining
equations. In what follows we drop the tilde's for brevity, and use the notation $S(t)$ for the bosonic collective spin
amplitude, $S(t) = \sigma_{gs}(t)/\sqrt{N}$, for compatibility with other papers.

To avoid exceeding the accuracy, we omit the spin transition frequency $\omega_{sg}$ and frequency
corrections $\delta_c$, $\delta_s(t)$ when they come in the sum with large Raman mismatches,
and neglect the terms of the order higher than 1 in $\g/|\Delta^{(m)}|\ll 1$ in the resulting equations.

Another important assumption is the large enough frequency mismatch 2$\omega_{sg}$ between the quantized field
frequencies supported by the Raman transitions in the memory and the luminescence channel,
$2|\omega_{sg}|\gg \g$, see figure 1.
This implies weak coupling of the memory to the luminescence channel, which is due to the Raman Stokes
transitions g - f - s in the far wing of this two-quantum transition spectral line. This makes it possible to consider the terms,
responsible for the interplay between the two channels, in the lowest (zero'th) approximation in
$\g/|\Delta^{(l)}|\ll 1$. We arrive at
 $$
\dot{\cal E}(t) = -\left[\kappa
+\frac{g^{(m)2} N{\g}}{\Delta^{(m)2}}\right]{\cal E}(t) +
\frac{ig^{(m)}\sqrt{N}}{\Delta^{(m)}}\left(1+\frac{i{\g}}{\Delta^{(m)}}\right)
\Omega^{(m)}(t){S}(t)+
 $$
 \begin{equation}
 \label{dot_E}
\frac{ig^{(l)}\sqrt{N}}{\Delta^{(l)}}
\Omega^{(l)}(t)e^{i2(\omega_{sg}t + \varphi_s(t))}{S}^{\dagger}(t)+ \sqrt{2\kappa}{\cal E}_{in}(t) +
 {F}_{\cal E}(t),
 \end{equation}
 $$
\dot{{S}}(t) = -{\g}\left(\frac{|\Omega^{(m)}(t)|^2}{\Delta^{(m)2}} +
\frac{|\Omega^{(l)}(t)|^2}{\Delta^{(l)2}}\right) S(t)+
\frac{ig^{(m)}\sqrt{N}}{\Delta^{(m)}}\left(1+\frac{i{\g}}{\Delta^{(m)}}\right)
\Omega^{(m)*}(t){\cal E}(t)+
 $$
 \begin{equation}
\label{dot_S}
\frac{ig^{(l)}\sqrt{N}}{\Delta^{(l)}}
\Omega^{(l)}(t)e^{i2(\omega_{sg}t+\varphi_s(t))}{\cal E}^{\dagger}(t)+{F}_{S}(t).
 \end{equation}
One can observe here the terms oscillating at the frequency $2(\omega_{sg} + \delta_s(t))$,
which represents the frequency detuning of the Raman Stokes
transitions  from resonance in the right $\Lambda$-scheme. This terms
are due to the luminescence channel and are able to introduce to the memory operation some degree
of squeezing and entanglement, as we demonstrate below. The ground state g excitation by control
field in the right $\Lambda$-scheme results in the additional relaxation of spin amplitude at rate
$\g|\Omega^{(l)}|^2/\Delta^{(l)2}$ in (\ref{dot_S}), where the factor $\sim\g/\Delta^{(l)2}$ represents
the density of the spectral line g - f far from the resonance.

The Langevin sources $F_{\cal E}$ and ${F}_{S}$ on the right-hand side of (\ref{dot_E}), (\ref{dot_S})
are linear combinations of the previously defined atomic sources. Applying the same approximations,
we obtain
 \begin{equation}
 \label{noise_E}
F_{\cal E}(t) =
\frac{g^{(m)}\sqrt{2\g N}}{\Delta^{(m)}}F_{ge}(t),
 \end{equation}
 \begin{equation}
 \label{noise_S}
F_S(t) = \Omega^{(m)*}(t)\frac{\sqrt{2\g}}{\Delta^{(m)}} F_{ge}(t) +
g^{(l)}{\cal E}^\dagger(t)\frac{\sqrt{2\g}}{\Delta^{(l)}}F_{gf}(t).
 \end{equation}
Considering a weak (as compared to the control field)  quantized signal, $g^{(l)2}\la\E^\dagger\E\ra \ll |\Omega^{(m)}|^2$,
we neglect the term $\sim F_{gf}$ in (\ref{noise_S}).
To be more specific, we restrict ourselves to the case when both $\Lambda$-schemes are based on the
same sets of superfine levels and assume $g^{(m)} = g^{(l)} = g$, $\Omega^{(m)}(t) = \Omega^{(l)}(t) = \Omega(t)$, and $\Delta^{(m)} = \Delta^{(l)} = \Delta$.

%%%%%%%%%%%%%%%%%%%%%%%%%%%%%%%%%%%%%%%%%%%%%%%%%
%%%%%%%%%%%%%%%%%%%%%%%%%%%%%%%%%%%%%%%%%%%%%%%%%
%%%%%%%%%%%%%%%%%%%%%%%%%%%%%%%%%%%%%%%%%%%%%%%%%
%%%%%%%%%%%%%%%%%%%%%%%%%%%%%%%%%%%%%%%%%%%%%%%%%

 \section{Spectral filtering}

In this section, we introduce the two-band representation for both the noise sources and the observables.
This allows us to take into account the side-band atomic noise sources on a par with the side-band quantized noise field,
entering the cavity.

The second adiabatic elimination is performed under the assumption that the frequency mismatch $2\omega_{sg}$
is much larger than all other frequency-like coefficients in (\ref{dot_E}), (\ref{dot_S}), that is,
much larger than the decay rates of the field
and the spin amplitudes, and the field-spin coupling due to the Raman transitions g - e - s and g - f - s. Given that
$|\omega_{sg}| \gg |\delta_s(t)|$, we introduce new slow amplitudes $\E^{(n)}$ and $S^{(n)}$, $n = m,l$,
which represent the observables evolution in the non-overlapping frequency bands associated with the memory
and the luminescence channels,
\begin{equation}
\label{two_bands}
{\cal E}(t)= {\cal E}^{(m)}(t)+ {\cal E}^{(l)}(t)e^{i2\omega_{sg}t}, \qquad
S(t)=S^{(m)}(t)+S^{(l)}(t)e^{i2\omega_{sg}t}.
\end{equation}
The similar representation is assumed for the noise operators $F_{n}(t)$, $n=\E,S$, and the input field $\E_{in}(t)$,
 \begin{equation}
\label{two_bands_noise}
{\cal E}_{in}(t) = {\cal E}^{(m)}_{in}(t) + {\cal E}^{(l)}_{in}(t) e^{i2\omega_{sg}t}, \qquad
F_n(t) = F^{(m)}_n(t)+F^{(l)}_n(t) e^{i2\omega_{sg}t}.
 \end{equation}
In order to define the latter introduced quantities, one has to perform spectral filtering of the initial noise sources (\ref{noise_E}, \ref{noise_S}), by multiplying their Fourier transforms by the non-overlapping filtering functions $\Pi^{(m)}(\omega)$,
centered at $\omega=0$, and $\Pi^{(l)}(\omega)$, centered at $\omega=-2\omega_{sg}$.
The filtering functions have the width $\sim |\omega_{sg}|$ and do not attenuate the Fourier amplitudes within
their width, hence, the correlation time of the filtered noise sources is of the order of $|\omega_{sg}|^{-1}$.
If the initial sources satisfy the relation
 \begin{equation}
 \label{comm_fast}
\left[F_n(t),F^\dagger_m(t')\right] =
A_{nm}(t)\,\delta(t - t'), \qquad n,m = \E,S,
 \end{equation}
where $A_{nm}$ is the noise covariance power, for the filtered sources we finally arrive at
 \begin{equation}
 \label{comm_m_l}
[F^{(i)}_n(t), F^{(j)\dagger}_m(t')] = \delta_{ij}
A_{nm}(t)\,\tilde\delta(t - t'), \quad i,j = m,l, \quad n,m = \E,S.
 \end{equation}
Here the delta-like function $\tilde\delta(t - t')$ has the temporal width $\sim |\omega_{sg}|^{-1}$.
The filtered noise amplitudes in two channels are mutually independent, are ``slow'' in terms of the first adiabatic
elimination, but can be viewed at as ``fast'' as compared to the observables $\E^{(i)}(t)$ and $S^{(i)}(t)$, $i = m,l$.
The same holds true for the input field.

The definitions (\ref{two_bands}) and (\ref{two_bands_noise}) are substituted into the evolution equations
(\ref{dot_E}) and (\ref{dot_S}). Omitting  fast oscillating terms, we arrive at the equations
for slow amplitudes $\E^{(i)}(t)$ and $S^{(i)}(t)$, $i = m,l$. The observables related to the
luminescence channel are  expressed in terms of these for the memory channel by means of
adiabatic elimination, that is $\dot\E^{(l)}$ and $\dot S^{(l)}$ are set
to zero as compared to $2\omega_{sg}\E^{(l)}$ and $2\omega_{sg} S^{(l)}$. This yields,
 \BE
 \label{dot_E_short}
\dot{\cal E}^{(m)}(t) =
-\left[\left(\kappa+\frac{g^2N{\g}}{\Delta^2}\right) + i\delta_R(t)\right]{\cal E}^{(m)}(t)
+\frac{ig\sqrt{N}}{\Delta}\left(1+\frac{i{\g}}{\Delta}\right)
\Omega(t)S^{(m)}(t) + \Phi_\E(t),
 \EE
 \BE
 \label{dot_S_short}
\dot{S}^{(m)}(t) = - \left[2\frac{\g |\Omega(t)|^2}{\Delta^2}  + i\delta_R(t)\right] S^{(m)}(t) +
\frac{ig\sqrt{N}}{\Delta}\left(1+\frac{i{\g}}{\Delta}\right)
\Omega^*(t){\cal E}^{(m)}(t) + \Phi_S(t).
 \EE
Here
 $$
 \delta_R(t) = - \frac{g^2N|\Omega(t)|^2}{2\omega_{sg}\Delta^2},
 $$
is the frequency correction induced by the Raman two-quantum transition in the right $\Lambda$-scheme. This frequency correction
is of the order of $|\delta_c\,\delta_s/\omega_{sg}| \ll |\delta_c|, |\delta_s|,$ and will be omitted due to our approximations.
The labeling of the observables of interest $\E^{(m)}$ and $S^{(m)}$ with ``$(m)$'' will be dropped for brevity.

The  combined Langevin noise operators in (\ref{dot_E_short}), (\ref{dot_S_short}) arise in the form
 \begin{equation}
\label{noise_E_comb}
\Phi_\E(t) =
-\frac{g\sqrt{N}}{2\omega_{sg}\Delta}\Omega(t){F}^{(l)\dagger}_{S}(t) +
{F}^{(m)}_{\cal E}(t) + \sqrt{2\kappa}{\cal E}^{(m)}_{in}(t),
 \end{equation}
\begin{equation}
\label{noise_S_comb}
\Phi_S(t) = -\frac{g\sqrt{N}}{2\omega_{sg}\Delta}\Omega(t)
\left[{F}^{(l)\dagger}_{\E}(t) + \sqrt{2\k}\E^{(l)\dagger}_{in}(t)\right] +
{F}^{(m)}_{S}(t).
\end{equation}
Coupling of the luminescence channel to the memory scheme leads to:

(i) An increase in the spin amplitude damping rate due to excitation of the initial state g by the
control field in the spectral wing of $g - f$ transition, see (\ref{dot_S}) and (\ref{dot_S_short}).
This has some impact on the memory efficiency through the excitations balance, see below.

(ii) An occurrence of new noise terms  $\sim F_n^{(l)\dagger}(t)$, $n=\E$, $S$, and
$\sim \E_{in}^{(l)\dagger}$ in (\ref{dot_E_short}), (\ref{dot_S_short}), which
are responsible for the creation of the field and spin quanta pairs, in analogy to many parametric
phenomena.  In these noise terms, the atomic noise is represented on a par with the
noise filed entering the cavity.  We demonstrate in the next sections that due to the process of four-wave mixing
these terms introduce additional noise to the memory readout signal, as well as some entanglement of the signal
with the spin subsystem.

%%%%%%%%%%%%%%%%%%%%%%%%%%%%%%%%%%%%%%%%%%%%%%%%%
%%%%%%%%%%%%%%%%%%%%%%%%%%%%%%%%%%%%%%%%%%%%%%%%%
%%%%%%%%%%%%%%%%%%%%%%%%%%%%%%%%%%%%%%%%%%%%%%%%%
%%%%%%%%%%%%%%%%%%%%%%%%%%%%%%%%%%%%%%%%%%%%%%%%%

 \section{Memory readout: quantum efficiency and noise }
 \label{sec_readout}

The output quantized field amplitude is given by the standard in-out relation,
 \begin{equation}
\label{in_out}
{\cal E}_{out}(t)=\sqrt{2\kappa}\,{\cal E}(t)-{\cal E}_{in}(t).
 \end{equation}
By the homodyne detection of the output signal on a time interval $[0, T]$, the observed quantity is
given by the projection of the signal on the normalized homodyne mode $\E^{(h)}(t) = \sqrt{2\k}e^{i\theta_h}\E_0(t)$,
 \BE
 \label{homo}
 \frac{ n_-}{\la  n \ra}  =  {\rm Re}\left(e^{-i\theta_h}\E_d\right), \qquad
\E_d = \sqrt{2\k}\int_0^T dt  \E_{out}(t)\E_0^*(t),
 \EE
where $n_-$ and $\la  n \ra$ are the difference and the average sum of counts in the arms of
detector, and
 $$
2\k \int_0^T dt |\E_0(t)|^2 = 1.
 $$
The commutation relation (\ref{noise_ini}) implies that the introduced amplitude $\E_d$ of the output signal temporal
mode is bosonic, $[\E_d,\E_d^\dagger]=1$. The directly measured quantity is an arbitrary quadrature component ${\rm Re}\left(e^{-i\theta_h}\E_d\right) \equiv Q_h$ of $\E_d$, and depends on the homodyne phase $\theta_h$.
In order to find the signal, we represent the solution of linear basic equations
(\ref{dot_E_short}, \ref{dot_S_short}) in terms of  dimensionless
Green functions $G_{nm} (t,t')$, $n, m = \E, S$,
 \BE
 \label{sol_E_gen}
{\E}(t) = G_{\E\E}(t,0){\E}(0) + G_{\E S}(t,0){S}(0)  +  \int_0^t dt' \sum_{n=\E,S}
G_{\E n}(t,t'){\Phi}_n(t'),
 \EE
 \BE
 \label{sol_S_gen}
S(t) = G_{S\E}(t,0){\E}(0) + G_{S S}(t,0){S}(0) + \int_0^t dt' \sum_{n=\E,S}
G_{Sn}(t,t'){\Phi}_n(t').
 \EE
Consider the memory retrieval. The starting spin amplitude $S(0)$ is most efficiently transferred to $\E_d$,
given $\E_{out}(t) \sim \sqrt{\eta}\cdot\sqrt{2\k}\E_0(t) S(0)$, when the projection (\ref{homo}) is
maximized. In view of (\ref{in_out}) and (\ref{sol_E_gen}), this is achieved when
 \BE
 \label{Green_es}
G_{\E S}(t,0) = \sqrt{\eta} e^{i\theta_{\!R}}\E_0(t),
 \EE
where $\eta \leq 1$ is quantum efficiency of the readout, and $\theta_R$ is an arbitrary phase shift.

Taking (\ref{Green_es}) for granted (see the next sections),  the observable is found to be
 $$
  \E_d =
\sqrt{\eta}e^{i\theta_{\! R}} S(0) +
2\k \int_0^{T}dt \E_0^*(t)
\left(G_{\E\E}(t,0)\E(0) - \frac{1}{\sqrt{2\k}}\E^{(m)}_{in}(t)\right) +
 $$
 \BE
 \label{observable_detailed}
2\k\int_0^{T}dt \E_0^*(t)
\int_0^{t} dt' \sum_{n=\E,S}G_{\E n}(t,t')\Phi_n(t') .
 \EE
Let us represent the general solution for the signal and spin amplitudes as
 \BE
 \label{sol_E_pm}
{\E}_d = G_{d\E}\E(0) + \sqrt{\eta}e^{i\theta_R}{S}(0) +  G_{d+}\Phi_d^{(+)} +
G_{d-}\Phi_d^{(-)},
 \EE
 \BE
 \label{sol_S_pm}
{S} =  G_{S\E}{\E}(0) +  G_{S S}{S}(0) + G_{S+}\Phi_S^{(+)} +
G_{S-}\Phi_S^{(-)},
 \EE
where we simplified the notation, $S(T)\to S$, $G_{S\E}(T,0)\to
G_{S\E}$, $G_{SS}(T,0)\to G_{SS}$.

The terms $\sim \Phi_d^{(+)}$, $\Phi_S^{(+)}$ include the positive-frequency (that is, the annihilation) noise
operators, which, as seen from (\ref{noise_E_comb}) and (\ref{noise_S_comb}), are associated with the memory channel.
The terms $\sim \Phi_d^{(-)}$, $\Phi_S^{(-)}$ are composed of the negative-frequency noise operators,
that are introduced by the luminescence channel. We assume that, by the definition,
 \BE
 \label{Phi_comm}
[\Phi_d^{(+)},\Phi_d^{(+)\dagger}] = [\Phi_d^{(-)\dagger},\Phi_d^{(-)}] =
[\Phi_S^{(+)},\Phi_S^{(+)\dagger}] = [\Phi_S^{(-)\dagger},\Phi_S^{(-)}] = 1.
 \EE
In order to preserve proper commutation relations for the bosonic amplitudes $\E_d$ and $S$,
the Green functions in (\ref{sol_E_pm}) and (\ref{sol_S_pm})
must obey the following relations,
 \BE
 \label{Ed_Ed}
[{\E}_d,{\E}_d^\dagger] = |G_{d\E}|^2 + \eta + |G_{d+}|^2 - |G_{d-}|^2 = 1,
 \EE
 \BE
 \label{S_S}
[{S},{S}^\dagger] = |G_{S\E}|^2 + |G_{SS}|^2 + |G_{S+}|^2 - |G_{S-}|^2 = 1,
 \EE
 \BE
 \label{Ed_S}
[{\E}_d,{S}^\dagger] = G_{d\E}G^*_{S\E} + \sqrt{\eta}e^{i\theta_R}G^*_{SS} +
G_{d+}G^*_{S+}[\Phi_d^{(+)},\Phi_S^{(+)\dagger}] -
G_{d-}G^*_{S-}[\Phi_S^{(-)\dagger},\Phi_d^{(-)}] = 0.
 \EE

By the memory readout, only the initial spin is assumed to be in a non-vacuum state. For the fluctuation of the observable
$\E_d$ this yields,
 $$
\Delta\E_d = \E_d - \la\E_d\ra = \sqrt{\eta}e^{i\theta_R}[S(0) - \la S(0)\ra] +
G_{d\E}\E(0) + G_{d+}\Phi_d^{(+)} + G_{d-}\Phi_d^{(-)}.
 $$
The uncertainty of an arbitrary quadrature amplitude of $\E_d$ is evaluated as
$\la(\Delta Q_d)^2\ra^{1/2}$, where $\Delta Q_d = {\rm Re}[e^{-i\theta_h}\Delta\E_d]$. By making use of
(\ref{Phi_comm}) and (\ref{Ed_Ed}), we arrive at
 \BE
 \label{quadr_d_noise_norm}
\la(\Delta Q_d)^2\ra =\frac{1}{4}\left\{1 + \eta\la : \left[ e^{i(\theta_R-\theta_h)}\Delta S(0) + h.c.\right]^2\! : \ra
+ 2|G_{d-}|^2 \right\},
 \EE
where ``$:\phantom{aa}:$'' denotes the normal ordering.

The fluctuation variance  (\ref{quadr_d_noise_norm}) is composed of the
contributions of (i) an excess over the vacuum level fluctuation of the relevant quadrature of the initial spin,
which is transferred to the output with quantum efficiency $\eta$, and (ii) the four-wave mixing noise due to the
presence of luminescence channel. By the retrieval of the initial spin in the vacuum state in absence of luminescence,
the output is also in the vacuum state, as it should be.

The added noise, which characterizes the memory device, is found after the removal of the retrieved
spin quadrature variance,
 \BE
 \label{quadr_d_noise_added}
\la(\Delta Q_d)^2\ra^{(add)} = \la(\Delta Q_d)^2\ra - \eta\la\left[{\rm Re}\left(e^{i(\theta_R-\theta_h)}\Delta S(0) \right)\right]^2\ra =
\frac{1}{4}\left\{1 - \eta  + 2|G_{d-}|^2 \right\}.
 \EE
This is our general result. Further, we will evaluate the impact of both the incomplete readout and the
four-wave mixing noise for the values of physical parameters typical for some experiments
using cells with alkaline atoms.

%%%%%%%%%%%%%%%%%%%%%%%%%%%%%%%%%%%%%%%%%%%%%%%%%
%%%%%%%%%%%%%%%%%%%%%%%%%%%%%%%%%%%%%%%%%%%%%%%%%
%%%%%%%%%%%%%%%%%%%%%%%%%%%%%%%%%%%%%%%%%%%%%%%%%
%%%%%%%%%%%%%%%%%%%%%%%%%%%%%%%%%%%%%%%%%%%%%%%%%

\section{Optimal control of the memory cell}
\label{sec_control}

Our goal is to evaluate the added noise (\ref{quadr_d_noise_added}) for a reasonable range of physical parameters of
the memory, that is, to find the readout quantum efficiency $\eta$, and to estimate the four-wave mixing noise contribution
$\sim |G_{d-}|^2 $.

The approaches, allowing for optimal memory control during the readout, such that the retrieved signal
has a predefined temporal shape and satisfies (\ref{Green_es}), were discussed in the literature
\cite{Gorshkov07,Stanojevic11,Dilley12}. Here we shall
use the version of impedance matching method, presented in \cite{Veselkova17b}, where the Raman memory
operation beyond the bad cavity limit (that is, for the output signals, whose duration is not arbitrary long as
compared to the cavity lifetime) was considered in details, including the non-stationary and relaxation
phenomena, and optimal phase matching of the signal and the control field.

Since the quantum efficiency $\eta$ arises as a parameter of the Green function (\ref{Green_es}), it can be found
by addressing a semiclassical version of the basic equations (\ref{dot_E_short}) and (\ref{dot_S_short}), and of the
in-out relation (\ref{in_out}), where the noise sources are dropped. The retrieved signal temporal mode ${\cal E}_0(t)$ is assumed to have a normalized quasi-Gaussian shape of duration $T$,
 \BE
 \label{E_0}
{\cal E}_0(t)=N_{\cal E}\left[\exp(-16(t/T-1/2)^2)-e^{-4}\right],\,\,
2\kappa\int_0^T dt {\cal E}^2_0(t)=1,
 \EE
where $N_{\cal E}$ is the normalization coefficient. The signal is truncated at the relative level $1/e^4 \sim 0.018$ and has the width at the relative level $\sim 1/e$, equal to $0.5$ of duration.
The ``inverse'' problem of estimating the control field time profile $\Omega(t)$ that matches the predefined time profile
of the retrieved signal beyond the bad cavity limit was considered in detail in \cite{Veselkova17b},
where the luminescence channel was not accounted for.
This channel introduces to the semiclassical equations only the additional decay rate of the spin amplitude, see
(\ref{dot_S}) and (\ref{dot_S_short}). This does not change the basic lines of the consideration given in \cite{Veselkova17b},
and we refer the reader to the cited paper. In brief, the main steps and issues arising are reduced to the following.

The time dependence of the spin excitations number that matches the needed time profile of the cavity field
is found by integrating the excitations balance,
 \BE
 \label{balance}
 \frac{d}{dt}\left(|\E_0|^2 + |S|^2\right) \approx  - 2\left(\kappa +
 \frac{g^2N{\g}}{\Delta^2}\right)|\E_0|^2  -
 4\frac{\g|\Omega S|^2}{\Delta^2},
 \EE
where $\Omega S$ is derived from (\ref{dot_E_short}),
 \BE
 \label{control}
 \Omega S \approx \frac{\Delta}{g\sqrt{N}}\left(1 - i\frac{\g}{\Delta}\right)
 \left[\frac{d}{dt}+\left(\kappa +
 \frac{g^2N{\g}}{\Delta^2}\right)\right]\E_0.
 \EE
Substituting the last expression to (\ref{dot_S_short}), one can finally calculate the spin phase,
 \BE
 \label{phi_S}
\phi_s(t) = -\frac{{\g}}{\Delta}
\int_0^t dt'\frac{1}{|S(t')|^2}\left[\frac{d}{dt'}+2\left(\kappa +
 \frac{g^2N{\g}}{\Delta^2}\right)\right]{\cal E}^2_0(t'),
 \EE
where $S(t) = |S(t)|e^{i\phi_s(t)}$.

Given the complex spin amplitude evolution is revealed, both the absolute value and the phase of the control field
are in turn found by making use of (\ref{control}).

\begin{figure}
 \begin{center}
 \includegraphics[width=0.4\linewidth]{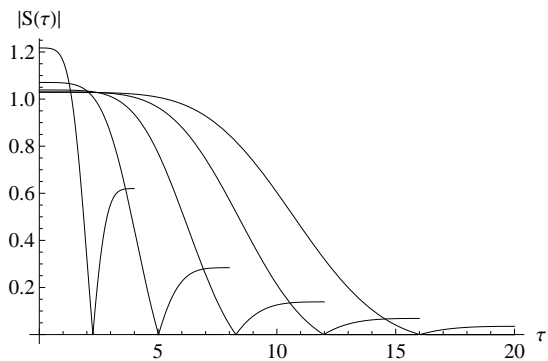}
 \caption{The spin amplitude for the normalized retrieved signal $\E_0(t)$
 of duration $2\k T$ =  4, 8, 12, 16, 20 (in the units of the cavity excitation lifetime).}
 \label{fig_2}
 \end{center}
 \end{figure}

An essential feature of the memory operation beyond the bad cavity limit,
revealed in \cite{Veselkova17b}, is that the back front of a signal of finite duration can be formed only
by means of a partial reabsorption of the field excitations by the atomic subsystem, as illustrated in figure \ref{fig_2}.
%%%%%%%%%%%%%%%%%%%%%%%%%%%%%%%%%%%%%%%%%%%%%%%%%
 %
 %
%%%%%%%%%%%%%%%%%%%%%%%%%%%%%%%%%%%%%%%%%%%%%%%%%
The reason for this is that the free decay of the cavity field after some time moment $t_s$ would
lead to the exponential form of the signal back front,
instead of that of $\E_0(t)$. This imposes limitations on the quantum efficiency and has some impact on
the phase properties of the system observables. An approach allowing to regularize the arising non-stationary
phase corrections was developed in \cite{Veselkova17b}.
In order to simplify evaluation of the four-wave mixing noise, we neglect here these  phase
corrections for the signal and the control field in some vicinity of $t_s$, and take (\ref{E_0}) for the signal shape.

The Green functions of the semiclassical version of (\ref{dot_E_short}) and (\ref{dot_S_short}) are
found by numerical integration, where we make use of the complex control field amplitude calculated
in the approach described above. Let us introduce the projections of the Green functions on the
signal temporal mode,
 \BE
 \label{projections}
P_{d\E}(T,t) = 2\kappa\int_{t}^T dt' \E_0^*(t') G_{\E\E}(t',t),  \qquad
P_{dS}(T,t) = 2\kappa\int_{t}^T dt' \E_0^*(t') G_{\E S}(t',t).
 \EE
This yields for the four-wave mixing contribution to the added noise variance (\ref{quadr_d_noise_added}),
  $$
|G_{d-}|^2=\frac{g^2 N}{(2\omega_{sg}\Delta)^2}
\int_0^{T} dt |\Omega(t)|^2\left\{
|P_{d\E}(T,t)|^2 \frac{2{\g}|\Omega(t)|^2}{\Delta^2} + \right.
 $$
 \BE
  \label{lum_power}
\left.\left[P_{d\E}(T,t)P_{dS}^*(T,t) \frac{2{\g} g\sqrt{N}\Omega(t)}{\Delta^2} +
{\rm c. c.}\right] +
|P_{dS}(T,t)|^2 \left(\frac{2{\g} g^2 N}{\Delta^2} +2\kappa\right)
\right\}.
 \EE
It is common to characterize the atom--field coupling with the cooperativity parameter
$C=g^2N/{\g}\kappa$ \cite{Gorshkov07}. For our numerical simulation, we assume the following values of the physical  parameters
corresponding to the off-resonant Raman regime: C = 200, ${\g}/2\pi$ = 3 MHz, $\kappa/2\pi$ = 2 MHz, $\Delta/2\pi$ = 200 MHz, $\omega_{sg}/2\pi$ =  10 MHz. The dimensionless time is measured in units of the cavity excitation lifetime $1/2\k$,
$\tau = 2\kappa t$, ${\cal T} = 2\kappa T$.

%%%%%%%%%%%%%%%%%%%%%%%%%%%%%%%%%%%%%%%%%%%%%%%%%
 %
 \begin{figure}
 \begin{center}
 \includegraphics[width=0.4\linewidth]{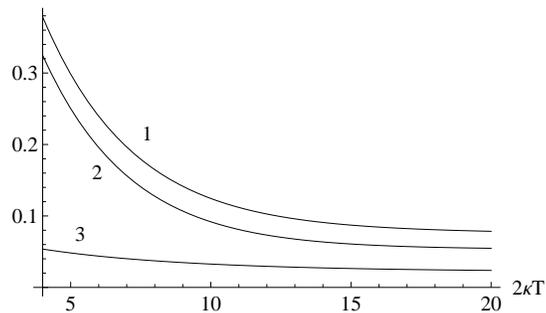}
 \caption{The variance $4\la(\Delta Q_d)^2\ra^{(add)}$  (curve 1)  and the contributions 1 - $\eta$ (2)
and $2|G_{d-}|^2$ (3), associated with the different from unity quantum efficiency and with the four-wave
mixing noise.}
 \label{fig_3}
 \end{center}
 \end{figure}
 %
%%%%%%%%%%%%%%%%%%%%%%%%%%%%%%%%%%%%%%%%%%%%%%%%%

We represent the retrieved signal noise (\ref{quadr_d_noise_added}) by plotting the variance $4\la(\Delta Q_d)^2\ra^{(add)}$, see curve 1 in figure \ref{fig_3}.
In order to reveal the role of non-adiabatic effects that arise by the memory operation beyond
the bad cavity limit, we present our results for a wide range of the signal duration, starting from the relatively
short pulses in the timescale of  $1/2\k$.

The vacuum noise contribution $(1 - \eta)$ appears in curve 2, where the memory readout quantum efficiency
was derived from the solution of the excitations balance equation (\ref{balance}) as $\eta = 1/|S(0)|^2$.
Note that in order to retrieve single excitation from the memory by $\eta<1$, the initial
number of spin excitations must exceed 1.

As we have demonstrated earlier \cite{Veselkova17a,Veselkova17b}, the quantum efficiency decrease
is basically due to the number of spin excitations $|S(\T)|^2$ retained in the memory by incomplete readout, and
to the field and spin relaxation terms  in (\ref{dot_E_short}, \ref{dot_S_short}), proportional to $\g$.
By the adopted values of the system parameters, just the steep increase in the number of unread excitations
for short signals is the main limiting factor for the memory quantum efficiency in the non-adiabatic regime.

The noise term $2|G_{d-}|^2$ introduced by the four-wave mixing is shown in curve 3. This noise contribution
does not demonstrate a comparably significant increase for short signals. An important feature
of this source of the memory imperfection is that for a large enough frequency mismatch $\omega_{sg}$ this term scales as $1/\omega_{sg}^2$, as follows from (\ref{lum_power}).

\section*{Acknowledgments}

This research was supported by the Russian Foundation for Basic Research (RFBR) under the projects 16-02-00180-a
and 18-02-00648-a. NIM acknowledges the RFBR grant for young researches 18-32-00255-mol-a.

%%%%%%%%%%%%%%%%%%%%%%%%%%%%%%%%%%%%%%%%%%%%%%%%%
%%%%%%%%%%%%%%%%%%%%%%%%%%%%%%%%%%%%%%%%%%%%%%%%%
%%%%%%%%%%%%%%%%%%%%%%%%%%%%%%%%%%%%%%%%%%%%%%%%%
%%%%%%%%%%%%%%%%%%%%%%%%%%%%%%%%%%%%%%%%%%%%%%%%%

 \appendix

 \section{Spin noise and entanglement}

In order to make our consideration more comprehensive, we briefly review here
the statistics of residual spin excitation $S(\T)=S$, as well as its entanglement with the retrieved signal.
Equation (\ref{sol_S_pm}) yields for the spin fluctuation,
 \BE
 \label{S_fluct_def}
\Delta S =
G_{SS}[S(0) - \la S(0)\ra] +  G_{S\E}\E(0) + G_{S+}\Phi_S^{(+)} + G_{S-}\Phi_S^{(-)}.
 \EE
The variance of an arbitrary spin fluctuation quadrature  $\Delta  Q_S = {\rm Re}( e^{-i\theta_S}\Delta S)$,
specified by the phase $\theta_S$, is
 \BE
 \label{quadr_S_noise_reduced}
\la(\Delta Q_S)^2\ra =\frac{1}{4}\left\{1 +  \la : \left[ e^{-i\theta_S}G_{SS}\Delta S(0) + h.c.\right]^2\! :\ra +
2|G_{S-}|^2 \right\},
 \EE
where we made use of (\ref{S_S}). The noise introduced by the spin quanta, created in pairs with the Raman
luminescence photons, is represented by the contribution  $\sim |G_{S-}|^2$ .

In terms of the signal--spin covariance matrix, the correlation between the two subsystems at $t=T$ is
described with
 $$
\frac{1}{2}\la\left(\Delta Q_d \Delta Q_S + \Delta Q_S \Delta Q_d\right)\ra =
\frac{1}{4}\left\{\la : \!\big[e^{-i(\theta_h-\theta_R)}\sqrt{\eta}\Delta S(0) + h.c.\big]
\big[ e^{-i\theta_S}G_{SS}\Delta S(0) + h.c.\big]\! : \ra +
\right.
 $$
 \BE
 \label{cross_norm}
\left.\left(e^{-i(\theta_h-\theta_s)} G_{d-}G^*_{S-}\la\Phi_S^{(-)\dagger}\Phi_d^{(-)}\ra + c.c.\right)\right\},
 \EE
where the commutation relation (\ref{Ed_S}) was used. The retrieved signal and the residual
spin are correlated (i) due to the partial transfer of the initial spin quadratures to both the signal and the
spin by an incomplete retrieval, and (ii) because of the parametric two-quantum
interaction similar to the $\chi^{(2)}$ nonlinearity in the luminescence channel.
Equation (\ref{cross_norm}) implies that for the vacuum initial state of the spin, the light-matter correlation is
completely of the parametric origin, as it should be.

%%%%%%%%%%%%%%%%%%%%%%%%%%%%%%%%%%%%%%%%%%%%%%%%%
%%%%%%%%%%%%%%%%%%%%%%%%%%%%%%%%%%%%%%%%%%%%%%%%%
%%%%%%%%%%%%%%%%%%%%%%%%%%%%%%%%%%%%%%%%%%%%%%%%%
%%%%%%%%%%%%%%%%%%%%%%%%%%%%%%%%%%%%%%%%%%%%%%%%%

\section{Self-consistency of the approach}
 \label{sec_sources}

It is instructive to reveal to which extent our basic equations (\ref{dot_E}, \ref{dot_S}) preserve bosonic
commutation relations of the observables. The macroscopic increments of the
relevant commutators are evaluated by making use of the observables increments of the form
 $$
 \Delta O(t) = A_O(t)\Delta t + \int\limits_{t}^{t+\Delta t}dt' F_O(t'),
 $$
where $O$ stands for $\E$ or $S$, the slow uniform terms in the right side of (\ref{dot_E}, \ref{dot_S})
are denoted by $A_O(t)$, and $F_O(t)$ are the noise sources (\ref{noise_E}, \ref{noise_S}).
Here the time increment $\Delta t$ is much shorter than the macroscopic evolution time but large
as compared to the noise correlation time. It is straightforward to demonstrate that given
$[\E(t),\E(t)^\dagger] = [S(t),S(t)^\dagger] =  1$, $[S(t),\E(t)^\dagger]=0$,  the macroscopic increments
of the commutators $[\E,\E^\dagger]$ and $[S,\E^\dagger]$ are equal zero, as it should be,
but for the increment of $[S,S^\dagger]$ we arrive at
 $$
\frac{\left<\Delta\left[S,S^\dagger\right](t)\right>}{\Delta t} =
-\frac{2{\g}|\Omega(t)|^2}{\Delta^2}.
 $$
Note that in our consideration the initial population of the ground state g is assumed to be
unchanged during the evolution. In the right side of the equation above stands the excitation rate of this state
by the off-resonant control field in the luminescence channel. Hence, the necessary condition for our theory
to be applicable is $(2\g|\Omega|^2/\Delta^2) T \ll 1$, when the relative decrease of the ground state population
is small.

%%%%%%%%%%%%%%%%%%%%%%%%%%%%%%%%%%%%%%%%%%%%%%%%%
%%%%%%%%%%%%%%%%%%%%%%%%%%%%%%%%%%%%%%%%%%%%%%%%%
%%%%%%%%%%%%%%%%%%%%%%%%%%%%%%%%%%%%%%%%%%%%%%%%%
%%%%%%%%%%%%%%%%%%%%%%%%%%%%%%%%%%%%%%%%%%%%%%%%%

 %\section{References}
 \bibliography{}
 \bibliographystyle{plain}

\end{document}